\documentstyle[aps,preprint]{revtex}

\newcommand{\be}{\begin{equation}}
\newcommand{\ee}{\end{equation}}
\newcommand{\Dlt}{\Delta}
\newcommand{\dlt}{\delta}
\newcommand{\om}{\omega}
\newcommand{\ep}{\varepsilon}

\newcommand{\br}{{\bf r}}
\newcommand{\bk}{{\bf k}}
\newcommand{\bS}{{\bf S}}
\newcommand{\bB}{{\bf B}}
\newcommand{\bt}{\beta}
\newcommand{\al}{\alpha}
\newcommand{\gm}{\gamma}
\newcommand{\Gm}{\Gamma}
\newcommand{\ra}{\rightarrow}

\newcommand{\lbd}{\lambda}
\newcommand{\prt}{\partial}

\tightenlines

\begin{document}

\draft

\title{Nonlinear spin relaxation in strongly nonequilibrium 
magnets}
\author{V.I. Yukalov} 

\address{Institut f\"ur Theoretische Physik, \\
Freie Universit\"at Berlin, Arnimallee 14, D-14195 Berlin, Germany}
\address{and}
\address{Bogolubov Laboratory of Theoretical Physics, \\
Joint Institute for Nuclear Research, Dubna 141980, Russia}

\maketitle

\begin{abstract}

A general theory is developed for describing the nonlinear relaxation 
of spin systems from a {\it strongly nonequilibrium} initial state, when, 
in addition, the sample is coupled to a resonator. Such processes are 
characterized by nonlinear stochastic differential equations. This makes
these strongly nonequilibrium processes principally different from the 
spin relaxation close to an equilibrium state, which is represented by 
linear differential equations. The consideration is based on a realistic 
microscopic Hamiltonian including the Zeeman terms, dipole interactions, 
exchange interactions, and a single-site anisotropy. The influence of 
cross correlations between several spin species is investigated. The 
critically important function of coupling between the spin system and a 
resonant electric circuit is emphasized. The role of all main relaxation 
rates is analyzed. The phenomenon of self-organization of transition 
coherence in spin motion, from the quantum chaotic stage of incoherent 
fluctuations, is thoroughly described. Local spin fluctuations are found 
to be the triggering cause for starting the spin relaxation from an 
incoherent nonequilibrium state. The basic regimes of collective 
coherent spin relaxation are studied.

\end{abstract}


\section{Introduction}

The problem of spin relaxation from a state close to equilibrium has a 
long history and is well studied, being related to the description of spin
motion in the vicinity of different magnetic resonances. This type of spin 
relaxation is usually characterized by linear differential equations, such 
as Bloch equations. The theory of spin motion close to equilibrium has 
been expounded in numerous literature, among which it would be possible to 
mention several good books [1--7].

Essentially nonlinear spin motion arises if the system is prepared in a 
strongly nonequilibrium initial state, e.g. with magnetization opposite to 
an external magnetic field, and, in addition, is coupled to a resonator. Such
nonlinear dynamics are commonly treated by the Bloch equations supplemented 
by the Kirchhoff  equation for a resonator electric circuit [8--11]. However, 
the phenomenological Bloch equations do not allow for the elucidation of 
different physical processes involved in the behaviour of the system and are 
not able to describe several, probably the most interesting, self-organized 
regimes of spin motion, as was demonstrated in Refs. [12--14]. Some physical 
models, based on microscopic spin Hamiltonians, have also been considered,
whose survey can be found in recent reviews [15,16]. But in each of these 
models one standardly studies only some particular substances and considers 
only a part of spin interactions, mainly secular dipole-dipole interactions, 
and one takes into account only some of the known attenuation processes. At
the same time, it is evident that taking care of only particular model 
elements can easily lead to wrong physical conclusions, since real physical 
materials always include several different characteristics competing with
each other. The study of nonlinear spin relaxation is of paramount importance
not solely owing to its theoretical beauty but also because it can be employed
in a variety of applications, such as the measurement of materials parameters,
ultrafast repolarization of solid-state targets, creation of sensitive field 
detectors, usage in quantum computing and others, as is discussed in reviews 
[15,16]. One of the major possible applications is in achieving the regime 
of superradiant operation by spin masers [13,17--19]. Punctuated nonlinear 
dynamics of spin assemblies can also be a new tool for information processing 
[20].

The aim of the present paper is to develop a general theory of nonlinear 
spin relaxation, being based on a realistic microscopic Hamiltonian including, 
in addition to the Zeeman terms, the main spin interactions, and taking 
account of the different major mechanisms of spin attenuation. By considering 
just some limited models, it is easy to come to false conclusions and to
predict fictitious physical effects that by no means can exist in real 
materials. It is only by carefully treating different competing mechanisms
that one can derive reliable physical implications.

\section{Basic Spin Hamiltonian}

Keeping in mind the applicability of the theory to a wide class of spin 
systems, we start with a rather general Hamiltonian including the major 
spin interactions the most often met in magnetic materials [1--7,21--23]. 
Let us consider a solid sample containing $N$ vector spins $\bS_i$ 
enumerated by the index $i=1,2,\ldots,N$. The spin operators $\bS_i$ can 
represent any particles of spin $S$, starting from $S=1/2$ to very high 
spin values. These can be nuclear or electronic spins, as in the standard 
problems of nuclear or electronic spin resonances [1--7,15]. Magnetic 
molecules, forming molecular magnets, can possess various spins ranging 
from $S=1/2$ up to $S=27/2$, as is reviewed in Refs. [16,19,24--26]. 
Bose-Einstein condensates of dilute gases (see reviews [27--30]), being 
placed in optical lattices can form localized clouds with an effective 
spin per site of order $10^2$ or $10^3$. Spin dynamics (mainly linear) is 
an intensively developing field of research, called spintronics [31].

The Hamiltonian  of a spin system can, generally, be separated into two 
parts,
\be
\label{1}
\hat H =\sum_i \hat H_i + \frac{1}{2}\; \sum_{i\neq j} \hat H_{ij} \; ,
\ee
the first term being related to individual spins, while the second 
representing spin interactions. The single-spin Hamiltonian 
\be
\label{2}
\hat H_i = -\mu_0\bB \cdot\bS_i - D(S_i^z)^2
\ee
consists of the Zeeman energy and the energy of the single-site magnetic 
anisotropy. Here $\mu_0\equiv\hbar\gm_S$, with $\gm_S$ being the gyromagnetic
ratio of a particle with spin $S$. For electronic spins, $\mu_0<0$, while
for nuclear spins $\mu_0$ can be either positive or negative. The total 
magnetic field
\be
\label{3}
\bB =  B_0{\bf e}_z + (B_1+H){\bf e}_x
\ee
contains external longitudinal, $B_0$, and transverse, $B_1$, magnetic 
fields, and also a feedback field $H$ of a resonator, if the sample is 
coupled to a resonant electric circuit. The anisotropy parameter $D$ is 
positive for an easy-axis anisotropy and negative in the case of an 
easy-plane anisotropy. 

The interaction Hamiltonian
\be
\label{4}
\hat H_{ij} =\sum_{\al\bt} D_{ij}^{\al\bt} S_i^\al S_j^\bt - 
J_{ij} \bS_i \cdot \bS_j 
\ee
includes dipole and exchange interactions. The dipolar tensor is
\be
\label{5}
D_{ij}^{\al\bt} = \frac{\mu_0^2}{r_{ij}^3}\; \left ( \dlt_{\al\bt} -
3 n_{ij}^\al\; n_{ij}^\bt \right ) \; ,
\ee
where $\al,\bt=x,y,z$ and
$$
r_{ij} \equiv |\br_{ij}| \; , \qquad 
{\bf n}_{ij} \equiv \frac{\br_{ij}}{r_{ij}} \; , \qquad 
\br_{ij}\equiv \br_i - \br_j \; .
$$
This tensor enjoys the properties
\be
\label{6}
\sum_\al D_{ij}^{\al\al} = 0 \; , \qquad
\sum_{j(\neq i)} D_{ij}^{\al\bt}  = 0\; ,
\ee
of which the first is exact and the second one is asymptotically exact for 
a macroscopic sample with a large number of spins $N\gg 1$. A positive 
exchange integral corresponds to ferromagnetic interactions and negative, 
to antiferromagnetic interactions.

It is convenient to represent the Hamiltonians through the ladder spin 
operators $S_i^\pm\equiv S_i^x \pm i S_i^y$. Then the single-spin term (2) 
writes as
\be
\label{7}
\hat H_i = -\mu_0 B_0 S_i^z - \; \frac{1}{2}\; \mu_0 (B_1 + H)\left (
S_i^+ + S_i^-\right ) - D \left ( S_i^z \right )^2 \; .
\ee
With the notation
\be
\label{8}
a_{ij} \equiv D_{ij}^{zz} \; , \qquad b_{ij}\equiv \frac{1}{4}
\left ( D_{ij}^{xx} - D_{ij}^{yy} - 2i D_{ij}^{xy} \right ) \; , \qquad
c_{ij} \equiv \frac{1}{2}\left ( D_{ij}^{xz} - i D_{ij}^{yz} \right ) \; ,
\ee
the interaction Hamiltonian (4) transforms to
$$
\hat H_{ij} = a_{ij}\left ( S_i^z S_j^z - \; \frac{1}{2}\; S_i^+ S_j^-
\right ) + b_{ij} S_i^+ S_j^+ + b_{ij}^* S_i^- S_j^- + 
$$
\be
\label{9}
+ 2c_{ij} S_i^+ S_j^z + 2 c_{ij}^* S_i^- S_j^z - J_{ij}\left ( 
S_i^+ S_j^- + S_i^z S_j^z \right ) \; .
\ee
The interaction parameters $a_{ij}=a_{ji}$, $b_{ij}=b_{ji}$, and 
$c_{ij}=c_{ji}$ are symmetric and have the property
\be
\label{10}
\sum_{j(\neq i)} a_{ij} = \sum_{j(\neq i)} b_{ij} = 
\sum_{j(\neq i)} c_{ij} = 0 \; ,
\ee
following from Eqs. (6).

The equations of motion for the spin operators are obtained from the 
Heisenberg equations and the commutation relations
$$
\left [ S_i^+,\; S_j^-\right ] = 2 \dlt_{ij} S_i^z \; , \qquad
\left [ S_i^z,\; S_j^\pm\right ] = \pm \dlt_{ij} S_i^\pm \; .
$$
In order to represent the evolution equations in a compact form, it is 
convenient to introduce the local fields
$$
\xi_0 \equiv \frac{1}{\hbar} \; \sum_{j(\neq i)} \left [ a_{ij} S_j^z +
c_{ij}^* S_j^- + c_{ij} S_j^+ + J_{ij}\left ( S_i^z - S_j^z\right )
\right ] \; ,
$$
\be
\label{11}
\xi \equiv \frac{i}{\hbar} \; \sum_{j(\neq i)} \left [ 2c_{ij} S_j^z -
\; \frac{1}{2}\; a_{ij} S_j^- + 2 b_{ij} S_j^+ + J_{ij}\left ( S_i^- -
S_j^-\right ) \right ]
\ee
and the effective force
\be
\label{12}
f \equiv -\; \frac{i}{\hbar} \; \mu_0 (B_1 + H) + \xi \; .
\ee
There is a characteristic frequency, the Zeeman frequency, which we denote 
as
\be
\label{13}
\om_0 \equiv - \; \frac{\mu_0}{\hbar}\; B_0 \; .
\ee
Then as the equations of motion for the spin operators, we obtain
\be
\label{14}
\frac{dS_i^-}{dt} = - i(\om_0 + \xi_0) S_i^- + fS_i^z + 
i\; \frac{D}{\hbar} \; \left ( S_i^- S_i^z + S_i^z S_i^- \right ) \; ,
\ee
with its Hermitian conjugate, and
\be
\label{15}
\frac{dS_i^z}{dt} = -\; \frac{1}{2}\; \left ( f^+ S_i^-  + S^+_i f
\right ) \; .
\ee
The following description of spin dynamics will be based on these 
equations.

\section{Triggering Spin Fluctuations}

Suppose that the spin system is prepared in a strongly nonequilibrium 
state, being polarized along the $z$-axis. What then could be the 
triggering mechanisms initiating spin motion and their relaxation to an 
equilibrium state? It is evident that imposing transverse magnetic fields 
would push the spins to move. But assume that there are no transverse 
magnetic fields at the initial time and no transverse coherence is imposed 
on the system. What then would initiate the spin motion? Here it is important 
to stress the role of local spin waves as of the triggering mechanism for 
starting the spin relaxation.

The appearance of spin waves is due to the local fields (11). In order to 
consider spin waves, or more generally, spin fluctuations that arise in a 
state which is not necessarily equilibrium, it is appropriate to work with 
the operator equations (14) and (15). Let us define the operator deviation
\be
\label{16}
\dlt S_i^\al \equiv S_i^\al\; -\; < S_i^\al>
\ee
from an average $<S_i^\al>$, which is not necessarily an equilibrium 
average, but which can be an average over a nonequilibrium statistical 
operator, though such that $<S_i^\al>$ weakly depends on the index $i$, 
because of which it can be taken out of the sums in Eqs. (11). Then, owing 
to Eqs. (10), we have
$$
\xi_0 =\frac{1}{\hbar}\; \sum_{j(\neq i)} \left [ a_{ij} \dlt S_j^z +
c_{ij}^* \dlt S_j^- + c_{ij}\dlt S_j^+ + J_{ij}\left ( \dlt S_i^z -
\dlt S_j^z\right ) \right ] \; ,
$$
\be
\label{17}
\xi = \frac{i}{\hbar} \; \sum_{j(\neq i)} \left [ 2 c_{ij}\dlt S_j^z - \;
\frac{1}{2}\; a_{ij}\dlt S_j^- + 2b_{ij}\dlt S_j^+ + J_{ij}\left (
\dlt S_i^- - \dlt S_j^-\right ) \right ] \; ,
\ee
which demonstrates that these local fields really correspond to local 
spin 
fluctuations.

To emphasize the role of the spin fluctuations, let us set $B_1=H=0$, 
that is, looking at the case when the transverse fields do not initiate 
the spin motion. And, respectively, let $<S_i^\pm>=0$, but the longitudinal
polarization be finite, $<S_i^z>\neq 0$. Then $S_i^\pm=\dlt S_i^\pm$. The
behaviour of spin fluctuations is characterized by linearizing Eqs. (14)
and (15) with respect to the operator deviations (16). The linearization 
of the single-site anisotropy term in Eq. (14) has to be done so that to 
satisfy the known exact relations for $S=1/2$ and $S\ra\infty$, which can 
be represented [16,32] as
\be
\label{18}
S_i^- S_i^z + S_i^z S_i^- = \left ( 2 -\; \frac{1}{S} \right )
<S_i^z>\; S_i^- \; .
\ee

Introduce the single-site anisotropy frequency
\be
\label{19}
\om_D \equiv (2S-1)\frac{D}{\hbar} 
\ee
and the effective spin frequency
\be
\label{20}
\om_s \equiv \om_0 - \om_D\; \frac{<S_i^z>}{S} \; ,
\ee
where $\om_0$ is defined in Eq. (13). Then, linearizing Eqs. (14) and 
(15), we find
\be
\label{21}
\frac{d}{dt}\; S_i^- = - i\om_s S_i^- + <S_i^z>\xi \; , \qquad
\frac{d}{dt}\; \dlt S_i^z = 0 \; .
\ee
The second of these equations, under the initial condition 
$\dlt S_i^z(0)=0$, gives $\dlt S_i^z=0$.

Now let us employ the Fourier transforms for the interactions
$$
a_{ij} = \frac{1}{N} \; \sum_k a_k e^{i\bk\cdot\br_{ij}} \; , \qquad
a_k = \sum_{j(\neq i)} a_{ij} e^{-i\bk\cdot\br_{ij}} \; ,
$$
with the analogous transforms for $b_{ij}$ and $J_{ij}$, and for the spin
operators
$$
S_j^\pm = \sum_k S_k^\pm e^{\mp i\bk\cdot\br_j} \; , \qquad
S_k^\pm = \frac{1}{N} \; \sum_j S_j^\pm e^{\pm i\bk\cdot\br_j} \; .
$$
Using the notation
\be
\label{22}
\al_k \equiv \om_s + \frac{1}{\hbar}\left ( \frac{a_k}{2} + 
J_k - J_0 \right ) <S_i^z> \; , 
\qquad \bt_k\equiv \frac{2}{\hbar}\; b_k\; <S_i^z> \; ,
\ee
from the first of Eqs. (14), we obtain
\be
\label{23}
\frac{d}{dt} \; S_k^- = -i\al_k S_k^- + i\bt_k S_k^+ \; .
\ee
Looking for the solution of the latter equation in the form
$$
S_k^- = u_k e^{-i\om_k t} + v_k^* e^{i\om_k t} \; ,
$$
we find the spectrum of spin waves
\be
\label{24}
\om_k = \sqrt{\al_k^2 -|\bt_k|^2} \; .
\ee
In the long-wave limit, one gets
\be
\label{25}
\om_k \simeq |\om_s| \left [ 1  - \; < S_i^z>\; \sum_{<j>} \;
\frac{a_{ij}+2J_{ij}}{4\hbar\om_s}\; (\bk\cdot\br_{ij})^2 \right ] \; ,
\ee
where $k\ra 0$, and the summation is over the nearest neighbours.

In this way, in the spin system there are always transverse fluctuations,
which can be named spin waves. The latter, as they have been described, 
are not necessarily the spin waves in an equilibrium state, as they are 
usually understood [33], but are to be considered in a generalized sense.
Under spin waves, we mean here just transverse spin fluctuations. It is
these transverse fluctuations that are responsible for triggering the 
initial motion of polarized spins, when there are no external transverse 
magnetic fields. This is why these transverse spin fluctuations can be 
called triggering spin waves. Taking into account such quantum spin 
fluctuations makes it possible to describe the dynamical regimes of spin 
motion, which do not exist for classical Bloch equations. And it becomes 
possible to develop a detailed picture of how the transverse spin 
coherence arises from initially chaotic fluctuations. This self-organized 
process of coherence emerging from chaos is one of the most interesting 
and challenging problems of spin dynamics.

\section{Spin Evolution Equations}

The equations of motion (14) and (15) for spin operators are highly 
nonlinear. The nonlinearity comes from two sources. One is caused by the
spin interactions accumulated in the local fluctuating fields (11). 
Another kind of nonlinearity enters through the effective force (12) 
containing feedback fields included in the term $H$. The treatment of the
nonlinear spin dynamics will be done here by means of the scale separation 
approach [11--15,34], which is a generalization of the averaging technique 
[35] to stochastic differential equations.

Notice, first of all, that there are two different spatial scales. One of 
them is related to local fields (11) describing random spin fluctuations 
(17), which is characterized by a spatial length of the order of the mean
interparticle distance $a_0$. At this length scale, chaotic quantum spin 
fluctuations prevail. Another length scale is the wavelength $\lbd\gg a_0$
corresponding to coherent effects associated with the characteristic spin 
rotation frequency $\om_s$. At the latter scale, coherent spin correlations
are important. These two different length scales allow us to distinguish 
two types of operators. One type are the local fluctuating fields (11), 
that is, the variables $\xi_0$, $\xi$, and $\xi^+$, and another type are 
the spin operators $S_i^-$, $S_i^+$, and $S_i^z$. The former, responsible
for local short-range fluctuations, can be represented by random variables
[2,5,11,16,36], while the latter keep track of long-range coherent effects.
Respectively, it is convenient to define two sorts of averaging with respect 
to the corresponding variables. Then the statistical averaging over spin 
operators will be denoted by the single angle brackets $<\ldots>$, while
the averaging over the random local fields will be denoted by the double 
angle brackets $\ll \ldots \gg$. The latter, treating the chaotic local 
spin fluctuations as white noise, are defined as
$$
\ll \xi_0(t)\gg \; =\; \ll \xi(t) \gg \; = 0 \; , \qquad
\ll \xi_0(t)\xi_0(t') \gg = 2\gm_3\dlt(t-t') \; ,
$$
\be
\label{26}
\ll \xi_0(t)\xi(t')\gg \; = \; \ll \xi(t)\xi(t')\gg \; = 0 \; ,
\qquad \ll \xi^*(t)\xi(t') \gg \; = 2\gm_3\dlt(t-t') \; ,
\ee
where $\gm_3$ is the width of inhomogeneous dynamic broadening.

It is worth stressing that the white-noise approximation (26) is not principal
and could be generalized to taking into account a coloured noise by including 
finite relaxation times. This, however, would result in much more complicated 
and cumbersome equations. It is therefore more convenient, following the ideas 
of the scale separation approach [11--15], to separate in the temporal
behaviour of spin correlations two parts, fast and slow. The fast part is
connected to the local spin fluctuations described by the spectrum of local 
spin waves (24). The characteristic frequencies of these fluctuations are 
defined by the near-neighbour spin coupling as well as by the applied external
 magnetic field. Here and in what follows, we assume that this external field
is sufficiently strong, so that the fluctuation spectrum (24) is characterized
by the frequencies of the order of the Zeeman frequency $\om_0$, which is
essentially larger than the frequency terms due to spin interactions. With
the time $2\pi/\om_0$ being the shortest among all other characteristic times,
the related fast spin fluctuations can be effectively treated as white noise, 
as is done in Eq. (26). The influence of spin correlations slowly decaying in 
time can be appropriately included into the transverse relaxation time $T_2$ 
determined by the strength of the spin-spin coupling allowing for dipolar as
well as exchange interactions. This effective relaxation time will also be 
taken into account in the following consideration, together with the effect
of line narrowing due to high spin polarization [6].

Averaging over spin operators, because of their long-range role, one can 
employ the decoupling
\be
\label{27}
<S_i^\al S_j^\bt>\; = \; < S_i^\al>< S_j^\bt> \qquad (i\neq j) \; .
\ee
Though this looks like a mean-field approximation, one should not 
forget that the restricted averaging, denoted by the single angle brackets 
$<\ldots>$, by definition, involves only the spin degrees of freedom, without 
touching the stochastic variables $\xi_0$ and $\xi^*$. Therefore the quantum 
fluctuations are not lost in decoupling (27) but are preserved because of the 
dependence of the spin averages $<S_i^\al>$ on the random variables $\xi_0$ 
and $\xi$. Then decoupling (27) is termed the stochastic mean-field
approximation [11--16].

A special care is to be taken in considering the single-site term of Eq. 
(14). When averaging the latter, one has to preserve the exact limiting
properties known for $S=1/2$ and $S\ra\infty$. The corresponding 
decoupling, correctly interpolating between the exact limiting behaviours
[16,19,32] is
\be
\label{28}
<S_i^- S_i^z + S_i^z S_i^-> \; = \left ( 2 - \; \frac{1}{S} \right )
<S_i^-><S_i^z> \; .
\ee
Thus, for $S=1/2$, expression (28) becomes zero, as it should be, and for
$S\ra\infty$, one has $2<S_i^-><S_i^z>$, again in agreement with the 
correct asymptotic behaviour.

Let us average the equations of motion (14) and (15) over the spin degrees 
of freedom, not touching the fluctuating random fields $\xi_0$ and $\xi$.
Our aim is to obtain the evolution equations for the following variables:
The {\it transition function}
\be
\label{29}
u \equiv \frac{1}{SN}\; \sum_{i=1}^N \; <S_i^-> \; ,
\ee
describing the average rotation of transverse spin components; the {\it 
coherence intensity}
\be
\label{30}
w \equiv \frac{1}{S^2N(N-1)} \; \sum_{i\neq j}^N \; <S_i^+ S_j^->\; ,
\ee
showing the level of coherence in the spin motion, and the {\it spin 
polarization}
\be
\label{31}
s \equiv \frac{1}{SN} \; \sum_{i=1}^N \; <S_i^z>\; ,
\ee
defining the average polarization per particle.

In order to have the evolution equations representing realistic spin 
systems, but not just some unreasonable models, an accurate account must 
be taken of the main relaxation mechanisms. Being based on unrealistic
models, omitting important existing attenuation processes, it would be 
easy to fall into the sin of predicting physical effects that in reality 
can never occur. We shall consider the following basic relaxation rates.

\vskip 2mm

(1) {\it Spin-lattice longitudinal attenuation} $\gm_1$, caused by 
spin-lattice interactions. The corresponding longitudinal relaxation time 
is $T_1\equiv 1/\gm_1$. For different materials, $\gm_1$ can be of different 
order. At low temperature, when spin-phonon interactions are suppressed, 
the parameter $\gm_1$ can be rather small. For instance, in polarized 
nuclear targets [16] at temperature of 1 K, one has $\gm_1\sim 10^{-5}$ 
s$^{-1}$. In molecular crystals below the blocking temperature of the 
order of 1 K, the spin-lattice rate can be between $\gm_1\sim 10^{-7}$ and
$10^{-5}$ s$^{-1}$ (see more details in Refs. [16,24--26]). Being small, 
this relaxation parameter may not play an essential role at the initial 
stage of spin motion, however, it always plays a principal role at the 
late stages of spin relaxation.

\vskip 2mm

(2) {\it Polarization pump rate} $\gm_1^*$, which is added to $\gm_1$ when 
the sample is subject to a permanent pump supporting a nonequilibrium level 
of the longitudinal spin polarization. This rate can be made  much larger 
than $\gm_1$. Thus, by means of dynamic nuclear polarization, the pump rate 
for nuclear spins in solids can be as large as $\gm_1^*\sim 0.01$ and $10$ 
s$^{-1}$ [16]. The sum of $\gm_1$ and $\gm_1^*$ will be denoted as 
\be
\label{32}
\Gm_1 \equiv \gm_1 + \gm_1^* \; .
\ee

\vskip 2mm

(3) {\it Spin dephasing rate} $\gm_2$, due to spin-spin interactions. This 
rate has been calculated by many authors, and the generally accepted value 
[1--7] writes as
\be
\label{33}
\gm_2 = n_0\rho \; \frac{\mu_0^2}{\hbar}\; \sqrt{S(S+1)} \; ,
\ee
where $\rho\equiv N/V$ is density and $n_0$ is a coefficient approximately
equal to the number of nearest neighbours. The process of spin dephasing 
is mainly due to dipolar forces. Exchange interactions slightly narrow the 
line width (33), yielding [4,6,21] a factor of about 0.8. The coefficient 
in Eq. (33) also depends on the type of lattice, so that the numerical factor 
here is approximate. The value of $\gm_2$ is usually larger than that of 
$\gm_1$. For example, in polarized solid targets [16] $\gm_2\sim 10^5$ 
s$^{-1}$, in molecular magnets [16,24] it is $\gm_2\sim 10^{10}$ s$^{-1}$. 
Inverse of $\gm_2$ defines the spin dephasing time $T_2\equiv 1/\gm_2$.

\vskip 2mm

(4) {\it Effective homogeneous broadening} $\gm_2(s)$ takes into account 
a correction to the spin dephasing rate $\gm_2$, appearing in the case of 
strongly polarized spin systems. Such a strong polarization can be achieved 
in magnetically ordered materials, by applying strong longitudinal magnetic 
fields, or by dynamic polarization techniques. This effective broadening
reads as
\be
\label{34}
\gm_2(s) = \gm_2(0)\left ( 1 - s^2 \right ) \; , \qquad
\gm_2(0)\equiv \gm_2 \; ,
\ee
where $s$ is an average spin polarization (31) and $\gm_2$ is given by Eq. 
(33). The derivation of Eq. (34) is explained in Appendix A. Under weak 
polarization, when $s^2\ll 1$, one has $\gm_2(s)\simeq\gm_2$.

\vskip 2mm

(5) {\it Static inhomogeneous broadening} $\gm_2^*$ is due to various 
magnetic defects, crystalline defects, field gradients, and a variety of 
additional interactions always present in any real materials [1--7,21,31]. 
Very often the inhomogeneity develops in matter not because of externally 
incorporated defects, but being due to the internal properties, when a 
heterogeneous state is more thermodynamically stable than a homogeneous 
state [37,38]. This, e.g., happens in many colossal-magnetoresistance 
materials [39--41] and in high-temperature superconductors [42--46], where 
there appears mesoscopic phase separation. In general, $\gm_2^*$ can be both 
smaller as well as larger than $\gm_2$. However in the majority of cases, to 
a very good approximation $\gm_2^*\sim\gm_2$. Summarizing the homogeneous and 
inhomogeneous mechanisms, discussed above, we denote the overall transverse 
relaxation rate as
\be
\label{35}
\Gm_2 \equiv \gm_2 \left ( 1 - s^2\right ) +\gm_2^* \; .
\ee

\vskip 2mm

(6) {\it Dynamic inhomogeneous broadening} $\gm_3$ is caused by fast dynamic 
spin fluctuations, or the local spin waves, discussed in Sec. III. It comes 
into play through the stochastic averaging (26). The value of the broadening, 
due to local spin waves, is of the order or smaller than $\gm_2$ [14--16,21]. 
As is emphasized in Sec. III, this dynamic broadening is crucially important 
at the initial stage of spin relaxation, when there are no applied transverse 
fields.

\vskip 2mm

(7) {\it Cross relaxation rates} arise when there are several spin species 
in the system. For example, if there are two types of spins, $S$ and $F$, 
then the dynamic broadening for spin $S$ becomes
\be
\label{36}
\gm_3 = \sqrt{\gm_{SS}^2 + \gm_{SF}^2} \; .
\ee
Cross correlations can influence other relaxation rates, especially if the 
Zeeman frequencies of the spins $S$ and $F$ are close to each other 
[1--7,15,16].

\vskip 2mm

(8) {\it Spin radiation rate} $\gm_r$ arises when there exist the so-called
wave packets of strongly correlated spins interacting with each other through 
the common radiation field. The possibility of the appearance of such an 
electromagnetic friction was, first, noticed by Ginzburg [47] and later 
discussed by many authors (see e.g. [48]). This collective radiation rate is
\be
\label{37}
\gm_r = \frac{2}{3\hbar}\; \rho\mu_0^2 S (kL_s)^3 \; ,
\ee
where $k$ is the wave vector of the radiating field and $L_s$ is an 
effective linear size of a spin packet radiating coherently. Rate (37) has 
earlier been obtained [47,48] in the classical approximation. In Appendix B, 
we briefly sketch how this rate can be derived in a fully quantum-mechanical 
picture. It is important to stress that the existence of rate (37) presupposes
the occurrence of monochromatic radiation with a well-defined constant spin 
frequency $\om_s$ and wave vector $k$, and that the radiation wavelength is 
much larger than the linear size $L_s$ of a spin packet, so that
\be
\label{38}
kL_s\ll 1 \qquad \left ( k\equiv \frac{\om_s}{c}\right ) \; .
\ee
If these conditions do not hold, no noticeable relaxation rate arises. And 
under the validity of these conditions, one has
\be
\label{39}
\frac{\gm_r}{\gm_2} \approx 0.1(kL_s)^3 \ll 1 \; .
\ee
The rate $\gm_r$ is so much smaller than $\gm_2$, and usually much smaller 
than $\gm_2^*$, that it can be safely neglected, being absolutely unable to
influence the motion of spins. Actually, Bloembergen [1] has already analysed 
this problem and come to the conclusion that the interaction of spins through
the magnetodipole radiation field is completely negligible. However, one may
put the following question. Suppose that the considered sample is ideally 
homogeneous, so that $\gm_2^*$ is very small, and let the initial spin 
polarization be very high, such that $s_0^2\approx 1$. Then the effective 
transverse rate (35) at the initial time $t=0$ can become rather small. 
Could then the radiation rate (37) play any noticeable role, at least at 
the very initial stage of spin motion? We study this problem below.

\vskip 2mm

(9) {\it Thermal noise attenuation} $\gm_T$ emerges when the spin system 
is coupled to a resonant electric circuit. The resonator Nyquist noise, due 
to the thermal fluctuations of current in the circuit creates a fluctuational 
magnetic field, which has to be included in the effective force (12). The 
magnitude of the thermal field, produced by the Nyquist noise, is well known 
[10]. It was found [12--16] that the resulting thermal attenuation is
\be
\label{40}
\gm_T = \frac{\eta\rho\mu_0^2\om}{4\hbar\gm N} \; 
{\rm coth}\; \frac{\om}{2\om_T} \; ,
\ee
where $\eta$ is a filling factor, $\om$ is the natural frequency of the 
electric circuit, $\gm$ is the resonator ringing width, and $\om_T\equiv
k_BT/\hbar$ is the thermal frequency. Bloembergen and Pound [8] first 
mentioned that, because of the macroscopic number of spins $N$ entering 
the denominator of $\gm_T$, the latter is unable to influence any spin 
motion in a macroscopic sample. This conclusion was confirmed by accurate 
calculations [12--16].

\vskip 2mm

(10) {\it Resonator relaxation rate} arises when the sample is coupled to 
a resonant electric circuit. Then in the effective force (12) the magnetic
field $H$ is the resonator feedback field. The role of this field will be 
thoroughly studied in what follows.

Summarizing all said above, for the spin averages (29) to (31), we obtain 
the evolution equations
\be
\label{41}
\frac{du}{dt} = - i (\om_s +\xi_0 -i\Gm_2) u + fs \; ,
\ee
\be
\label{42}
\frac{dw}{dt} = - 2\Gm_2 w + \left ( u^* f + f^* u\right ) s \; ,
\ee
\be
\label{43}
\frac{ds}{dt} = -\; \frac{1}{2}\left ( u^* f + f^* u \right ) -
\Gm_1 (s -\zeta) \; ,
\ee
supplemented by the initial conditions
$$
u(0) =u_0 \; , \qquad w(0)=w_0\; , \qquad s(0)=s_0 \; .
$$
In these equations, $\zeta$ is a stationary spin polarization, the 
characteristic spin frequency is
\be
\label{44}
\om_s = \om_0 - \om_D s \; ,
\ee
with $\om_0$ given by Eq. (13) and $\om_D$, by Eq. (19). The total 
longitudinal rate $\Gm_1$ is defined in Eq. (32) and the total transverse
rate $\Gm_2$, in Eq. (35). The effective force is
\be
\label{45}
f = -\; \frac{i}{\hbar}\; \mu_0 (B_1+H) +\xi +\gm_r u \; ,
\ee
where the last term is the friction force due to the interaction through
magnetodipole radiation, and $\gm_r$ is the magnetodipole radiation rate 
(37). Equations (41) to (43) are stochastic differential equations, since 
they contain the random variables $\xi_0$ and $\xi$, whose stochastic 
averages are given in Eqs. (26). The external transverse field $B_1$ and 
the resonator feedback field $H$ need yet to be specified.

\section{Resonator Feedback Field}

The resonator feedback field $H$ is created by the electric current of the 
coil surrounding the spin sample. We assume that the coil axis is along 
the axis $x$. The electric circuit is characterized by resistance $R$, 
inductance $L$, and capacity $C$. The spin sample is inserted into a coil 
of $n$ turns, length $l$, cross-section area $A_c$, and volume $V_c=A_cl$.
The electric current in the circuit is described by the Kirchhoff equation
\be
\label{46}
L\; \frac{dj}{dt} + Rj + \frac{1}{C} \; \int_0^t j(t')\; dt' = E_f - \;
\frac{d\Phi}{dt} \; ,
\ee
in which $E_f$ is an electromotive force, if any, and the magnetic flux
\be
\label{47}
\Phi = \frac{4\pi}{c}\; nA_c\eta m_x \; ,
\ee
where $\eta\approx V/V_c$ is a filling factor, is formed by the $x$-component
of the magnetization density
\be
\label{48}
m_x \equiv \frac{\mu_0}{V} \; \sum_i <S_i^x> \; .
\ee
The electric current, circulating over the coil, creates a magnetic field
\be
\label{49}
H =\frac{4\pi n}{cl}\; j \; .
\ee
The circuit natural frequency is
\be
\label{50}
\om \equiv \frac{1}{\sqrt{LC}} \qquad \left ( L \equiv 
4\pi\; \frac{n^2 A_c}{c^2l} \right )
\ee
and the circuit damping is
\be
\label{51}
\gm \equiv \frac{1}{\tau} = \frac{R}{2L} = \frac{\om}{2Q} \; ,
\ee
where $\tau$ is called the circuit ringing time and $Q\equiv\om L/R$ is 
the quality factor. Also, let us define the reduced electromotive force
\be
\label{52}
e_f \equiv \frac{cE_f}{nA_c\gm} \; .
\ee
Then the Kirchhoff equation (46) can be transformed to the equation
\be
\label{53}
\frac{dH}{dt}  + 2\gm H + \om^2 \int_0^t H(t')\; dt' =
\gm e_f - 4\pi\eta\; \frac{dm_x}{dt}
\ee
for the feedback magnetic field created by the coil.

The feedback equation (53) can be represented in another equivalent form 
that proved to be very convenient for defining the feedback field [12--15].
For this purpose, we involve the method of Laplace transforms and 
introduce the transfer function
\be
\label{54}
G(t) =\left ( \cos\om' t -\; \frac{\gm}{\om'}\; \sin\om' t \right )
e^{-\gm t} \; , 
\ee
where
$$
\om' \equiv\sqrt{\om^2-\gm^2} \; .
$$
Thus, we transform the feedback-field equation (53) to the integral 
representation
\be
\label{55}
H = \int_0^t G(t-t') \left [ \gm e_f(t') - 4\pi\eta\dot{m}_x(t')
\right ] \; dt'\; ,
\ee
in which
\be
\label{56}
\dot{m}_x(t) \equiv \frac{1}{2}\; \rho\mu_0 S \; 
\frac{d}{dt}\; (u^* + u) \; .
\ee

Let the resonant part of the reduced electromotive force (52) be
\be
\label{57}
e_f(t) = h_2\cos\om t \; .
\ee
And let us introduce the notation
\be
\label{58}
\nu_2 \equiv \frac{\mu_0 h_2}{2\hbar} \; .
\ee

As usual, we assume that all attenuation parameters are much smaller than 
the characteristic spin frequency $\om_s$. Then Eq. (55) can be solved by 
an iteration procedure, which in first order gives
\be
\label{59}
\frac{\mu_0H}{\hbar} = i (\al u - \al^*u^*) + 2\bt\cos\om t \; .
\ee
Here the coupling function
\be
\label{60}
\al =\gm_0 \om_s \left [ 
\frac{1-\exp\{-i(\om-\om_s)t-\gm t\} }{\gm+i(\om-\om_s)} +
\frac{1-\exp\{-i(\om+\om_s)t-\gm t\} }{\gm-i(\om+\om_s)} \right ]
\ee
describes the coupling of spins with the resonator and the function
\be
\label{61}
\bt = \frac{\nu_2}{2}\left ( 1 - e^{-\gm t} \right )
\ee
characterizes the action of the resonator electromotive force on spins. In 
Eq. (60) the notation for the natural spin width
\be
\label{62}
\gm_0 \equiv \frac{\pi}{\hbar}\; \eta\rho\mu_0^2 S
\ee
is employed.

The spin-resonator coupling can be characterized by the dimensionless 
coupling parameter
\be
\label{63}
g \equiv \frac{\gm\gm_0\om_s}{\gm_2(\gm^2+\Dlt^2)} \; ,
\ee
in which $\Dlt\equiv\om-|\om_s|$ is the detuning. As is evident from Eq. 
(60), an efficient spin-resonator coupling is possible only when the 
detuning from the resonance is small, such that
\be
\label{64}
\frac{|\Dlt|}{\om} \ll 1 \qquad (\Dlt\equiv\om-|\om_s|) \; .
\ee
When the resonance is sufficiently sharp, so that $|\Dlt|<\gm$, then the 
coupling function (60) reduces to
\be
\label{65}
\al = g\gm_2\left ( 1  - e^{-\gm t} \right ) \; .
\ee
Thus, the resonator feedback field $H$ is defined by Eq. (59), in which 
$\al$ is given by Eq. (65) and $\bt$, by Eq. (61).

\section{Averaged Evolution Equations}

The resonator field, defined in Eq. (59), has to be substituted in the 
effective force (45) entering the evolution equations (41) to (43). In Eq. 
(45), we also need to specify the external magnetic field $B_1$. In general,
the latter may contain a constant part and an alternating term. So, let us 
take this transverse field in the form
\be
\label{66}
B_1 = h_0 + h_1\cos\om t \; .
\ee
In what follows, we shall use the notation
\be
\label{67}
\nu_0 \equiv \frac{\mu_0 h_0}{\hbar} \; , \qquad 
\nu_1 \equiv \frac{\mu_0 h_1}{2\hbar} \; .
\ee
Equations (41) to (43) are stochastic differential equations, containing 
the random variables $\xi_0$ and $\xi$ describing local spin fluctuations. 
In order to derive the evolution equations in terms of ordinary differential 
equations, we have to accomplish the averaging over random fluctuations. 
This can be done by following the scale separation approach [11--16], the 
usage of the stochastic averages (26), and by invoking the known techniques 
of treating stochastic variables [49].

Keeping in mind that the attenuation parameters are substantially smaller 
than the characteristic spin frequency $\om_s$, we notice from Eqs. (41) 
to (43) that the function $u$ can be classified as fast, being compared 
with the temporal behaviour of the functions $w$ and $s$. The latter play 
the role of temporal quasi-invariants with respect to $u$.

Fist, we substitute into Eqs. (41) to (43) the effective force (45), the 
resonator field (59), and the transverse magnetic field (66). This results 
in the equations
\be
\label{68}
\frac{du}{dt} = - i(\om_s+\xi_0)u - (\Gm_2 -\al s -\gm_r s)u + f_1 s - 
\al s u^* \; ,
\ee
\be
\label{69}
\frac{dw}{dt} = -2(\Gm_2 - \al s - \gm_r s) w + \left ( u^* f_1 + 
f_1^* u \right ) s - \al s \left ( u^2 +(u^*)^2\right ) \; ,
\ee
\be
\label{70}
\frac{ds}{dt} = -(\al+\gm_r)w -\; \frac{1}{2}\left ( u^* f_1 +
f_1^* u \right ) - \Gm_1(s-\zeta) + \frac{1}{2}\;\al \left (
u^2 + (u^*)^2 \right ) \; ,
\ee
in which
\be
\label{71}
f_1 \equiv -i\nu_0 - 2i (\nu_1+\bt)\cos\om t + \xi \; .
\ee
Then we solve Eq. (68) for the fast variable $u$, keeping the 
quasi-invariants fixed, which yields
$$
u=u_0\exp\left \{ - (i\om_s +\Gm_2 -\al s -\gm_r s) t - 
i \int_0^t \xi_0(t')\; dt' \right \} +
$$
\be
\label{72}
+ s \int_0^t f_1(t')\exp\left\{ - (i\om_s +\Gm_2 -\al s -\gm_r s) (t -t')
- i \int_{t'}^t \xi_0(t'')\; dt'' \right \}\; dt' \; .
\ee
Solution (72) must be substituted in Eqs. (69) and (70) for the slow 
functions $w$ and $s$. After this, the latter equations have to be averaged 
over time and over the stochastic variables $\xi_0$ and $\xi$, again keeping 
the quasi-invariants fixed. To slightly simplify the resulting equations, 
one can take the initial condition for the transition function $u$ in the 
real form, such that $u_0^*=u_0$, which is not principal but just makes 
the equations less cumbersome.

To present the resulting equations in a compact form, we introduce the 
{\it effective attenuation}
\be
\label{73}
\Gm_3 \equiv \gm_3 + \frac{\nu_0^2\Gm}{\om_s^2+\Gm^2} \; - \;
\frac{\nu_0(\nu_1+\bt)\Gm}{\om_s^2+\Gm^2}\; e^{-\Gm t} +
\frac{(\nu_1+\bt)^2\Gm}{\Dlt^2+\Gm^2}\left ( 1  - e^{-\Gm t}\right ) \; ,
\ee
in which
\be
\label{74}
\Gm\equiv \Gm_2 +\gm_3 - (\al+\gm_r) s \; .
\ee
And finally, after the described averaging, we obtain the evolution 
equations
\be
\label{75}
\frac{dw}{dt} = - 2(\Gm_2 -\al s -\gm_r s) w + 2\Gm_3 s^2 \; ,
\ee
\be
\label{76}
\frac{ds}{dt} = - (\al+\gm_r) w - \Gm_3 s - \Gm_1(s-\zeta) \; .
\ee

These equations are very general. They include various attenuation processes,
described in Sec. IV, and take into account transverse constant and 
alternating fields (66), as well as the resonator electromotive force (57) 
entering through function (61). The resonator feedback field is responsible 
for the appearance of the coupling function (65). Notice that the radiation
relaxation rate $\gm_r$, defined in Eq. (37), enters everywhere together 
with the spin-resonator coupling $\al$. However their values are 
drastically different. Since
$$
\frac{\gm_r}{\al} \sim 0.1 \frac{\gm}{\om_s}\; (kL_s)^3 \ll 1 \; ,
$$
the value of $\gm_r$ is so incomparably smaller than $\al\sim g\gm_2$, 
that it is evident, in the presence of a resonator, the rate $\gm_r$ must 
be forgotten.

Moreover, even when there is no resonator, so that $\al=\bt=0$, the 
radiation rate $\gm_r$ plays no role, since it is much smaller than $\gm_2$,
$\gm_2^*$, and $\gm_3$. One might think that $\gm_r$ could play a role in 
the following unrealistic case. Let us imagine an absolutely ideal lattice 
with no inhomogeneous broadening, that is, let us set $\gm_2^*=0$, which 
is certainly a purely imaginary situation. Then, according to Eq. (35), 
one has $\Gm_2=\gm_2(1-s^2)$. Assume that the spin system is completely 
polarized, with $s_0=1$. Hence, at the initial time, $\Gm_2=0$. Could then 
the spin motion be started by the term with $\gm_r$? The answer is evident: 
As far as the largest terms in both Eqs. (75) and (76) are those 
containing $\Gm_3$, the terms with $\gm_r$ are always negligible, even if 
$\Gm_2=0$. Even more, functions (30) and (31), by their definition, satisfy 
the inequality
\be
\label{77}
w + s^2 \leq 1 \; .
\ee
Therefore, if one sets $s_0=1$, then $w_0=0$, and the term $\gm_r w$ 
simply disappears from the equations. Vice versa, if one sets a noticeable 
$w_0\sim 1$, then $s^2\ll 1$, and $\Gm_2\approx\gm_2\gg\gm_r$. In this 
way, the radiation rate $\gm_r$ never plays any role in the spin motion, 
which is in agreement with the estimates by Bloembergen [1].

Note that the situation in spin systems is principally different from that 
happening in atomic systems. In the latter, both the linewidth 
$\gm_2=2|{\bf d}|^2k^3/3$ as well as the collective radiation rate 
$\gm_r=(2/3)|{\bf d}|^2k^3N_c$, where $N_c$ is the number of correlated 
atoms, forming a wave packet, are caused by the same physical process, by 
the interaction of atoms with their radiation field. Hence 
$\gm_r/\gm_2=N_c\gg 1$, which results in the coherentization of the dipole 
transitions. This is possible even if $kL\gg 1$, but the number of atoms 
in a partial wave packet is $N_c\gg 1$, since $\gm_r/\gm_2=N_c\gg 1$. 
Contrary to this, in spin systems the linewidth $\gm_2$, given in Eq. (33), 
is due to direct dipole-dipole interactions, while the radiation rate (37) 
is a result of the spin interactions with their radiation field. This is 
why in the latter case, one always has $\gm_r\ll\gm_2$, and the radiation 
rate $\gm_r$ plays no part in the motion of spins.

We may also notice that in the effective attenuation (73) the terms due to 
the presence of a constant transverse field are less important than the terms 
caused by the local spin fluctuations and by the alternating transverse 
fields. Therefore, omitting the terms corresponding to a permanent 
transverse magnetic field, we have
\be
\label{78}
\Gm_3 = \gm_3 + \frac{(\nu_1+\bt)^2\Gm}{\Gm^2+\Dlt^2}\; \left ( 1 -
e^{-\Gm t}\right ) \; .
\ee
Finally, we obtain the evolution equations
\be
\label{79}
\frac{dw}{dt} = - 2(\Gm_2 -\al s) w + 2\Gm_3 s^2 \; ,
\ee
\be
\label{80}
\frac{ds}{dt} = -\al w - \Gm_3 s - \Gm_1(s-\zeta) \; ,
\ee
describing the averaged motion of spins.

\section {Coherence Emerging from Chaos}

One of the most intriguing questions is how the spin motion could become 
coherent if initially it was not. This is a particular case of the general 
physical problem of how coherence emerges from chaos.

Being interested in a self-organized process of arising coherence, let us 
consider the case, when there are no external transverse fields pushing 
spins, that is $\nu_1=\bt=0$. Then Eq. (78) yields $\Gm_3=\gm_3$. Assume
also that there is no pumping, so that $\gm_1^*=0$, hence $\Gm_1=\gm_1$. 
Under these conditions, the initial spin motion, for the time $t$ such 
that
\be
\label{81}
\gm_1 t \ll 1 \; , \qquad \gm_2 t \ll 1 \; , \qquad \gm_3 t\ll 1 \; ,
\ee
follows from Eqs. (79) and (80) in the form
$$
w\simeq w_0 +2 \left [ \gm_3 s_0^2 -\gm_2 \left ( 1 - s_0^2 + \kappa
\right ) w_0 \right ]\; t \; ,
$$
\be
\label{82}
s \simeq s_0 - [ ( \gm_1 +\gm_3) s_0 - \gm_1 \zeta ] \; t \; ,
\ee
where the inhomogeneity coefficient is introduced,
\be
\label{83}
\kappa \equiv \gm_2^* /\gm_2 \; .
\ee
If at the initial time no transverse polarization is imposed on the system,
and the initial coherence function is zero, $w_0=0$, nevertheless the 
coherent spin motion starts developing according to the law
\be
\label{84}
w \simeq 2\gm_3 s_0^2 t \qquad (w_0 = 0) \; ,
\ee
provided there is an initial longitudinal polarization $s_0\neq 0$. The 
initiation of the emerging coherent motion is caused by local spin 
fluctuations creating the effective rate $\gm_3$. Recall that in the Bloch 
equations coherent motion never appears if it is not imposed by the initial 
conditions. Contrary to this, Eqs. (79) and (80) take into account the local 
spin fluctuations triggering the motion of spins. The second of Eqs. (82), 
keeping in mind that usually $\gm_1\ll\gm_3\sim\gm_2$, can be simplified 
to
\be
\label{85}
s\simeq s_0 ( 1  -\gm_3 t) \; .
\ee

At the initial stage of spin motion, their coherence is yet incipient, 
and the motion is mainly governed by quantum chaotic spin fluctuations. 
The coherentization of the transverse motion goes through the resonator 
feedback field and the growing coupling function (65). The quantitative 
change in the spin motion happens when the coupling function (65) becomes 
so large that the term $(\Gm_2-\al s)$ in Eq. (79) goes negative, which 
means that an efficient generation of coherence has started in the system. 
This is analogous to the beginning of maser generation [15--19]. The 
moment of time, when the regime of mainly chaotic quantum fluctuations 
transforms into the regime of predominantly coherent spin motion, can be 
called the {\it chaos time}. This time $t_c$ is defined by the equality 
$\al s=\Gm_2$, that is by the equation
\be
\label{86}
\al s = \gm_2 (1 -s^2) + \gm_2^* \qquad (t=t_c) \; .
\ee
From here, the estimate for the chaos time is
\be
\label{87}
t_c = \tau \ln\; \frac{gs_0}{gs_0-1+s_0^2-\kappa} \; ,
\ee
where $\tau$ is the resonator ringing time defined in Eq. (51). The regime 
of chaotic spin fluctuations lasts till the chaos time (87), after which 
the coherent stage of spin motion comes into play. As is clear from the 
above equations, the transformation from the chaotic to coherent regime 
goes as a gradual crossover. Notice that the quantity $1-s_0^2+\kappa$ is 
positive since $s_0^2\leq 1$. Then, in order that the chaos time (87) be 
positive and finite, the inequality
\be
\label{88}
gs_0 > 1-s_0^2 +\kappa > 0
\ee
must hold. For a strong spin-resonator coupling, when $gs_0\gg 1$, the 
chaos time (87) reduces to
\be
\label{89}
t_c \simeq \frac{\tau}{gs_0}\; \left ( 1 - s_0^2 +\kappa\right ) \; .
\ee

As is seen, there exists a well defined stage of chaotic spin fluctuations, 
with a finite chaos time $t_c>0$, after which the coherent regime develops,
if $gs_0>0$. The coupling parameter $g$ is defined in Eq. (63), from which 
it follows that one should have $\om_s s_0>0$. Assuming that the initial
spin polarization is positive, $s_0>0$, one gets the requirement that 
$\om_s>0$. The latter, by definition (44), is equivalent to the condition 
$\om_0>\om_Ds$. Moreover, the coupling function (65) is obtained under the 
resonance condition (64), which implies that $\om_s$ has to be close to 
the resonator natural frequency $\om$. There are two ways of preserving 
the resonance condition (64). First, one can impose a sufficiently strong 
external magnetic field $B_0$, such that the frequency $\om_0$, given by 
Eq. (13), would be much larger than $\om_D$, defined in Eq. (19). This 
becomes trivial for $S=1/2$, when $\om_D=0$. If $\om_0\gg\om_D$, then it 
is easy to realize the resonance condition (64), with $\om_s\approx\om$ 
and slightly varying in time detuning $\Dlt=\om-\om_s$.

The second way of keeping the resonance condition (64) is by means of the 
chirping effect [16,19]. This requires to vary in time the external 
magnetic field $B_0$ so that to maintain the equality
\be
\label{90}
\frac{\mu_0B_0}{\hbar} + (\om + \om_D s) = \Dlt \; ,
\ee
with a fixed detuning.

\section{Coherent Spin Relaxation}

After the chaos time (87), the motion of spins becomes more and more 
coherent, being collectivized by the resonator feedback field, with the 
coupling function $\al$ reaching the value $g\gm_2$. At the transient 
stage, when $t>t_c$ but $t\ll T_1$, we may neglect the term with $\gm_1$ 
in Eq. (80). Assuming that there is no pumping, that is $\gm_1^*=0$, one 
has $\Gm_1=\gm_1$. Let us continue studying the case of the self-organized 
coherent spin motion, when there are no transverse external fields, so that 
$\nu_1=\bt=0$, hence $\Gm_3=\gm_3$. When the coherence is well developed, 
then the main term in Eq. (79) is the first one, while the term with $\gm_3$ 
can be neglected. Under these conditions, and using expression (35) for the 
rate $\Gm_2$, Eqs. (79) and (80) reduce to the form
\be
\label{91}
\frac{dw}{dt} = - 2\gm_2 \left ( 1 - s^2 +\kappa -gs\right ) \; w \; ,
\ee
\be
\label{92}
\frac{ds}{dt} = - g\gm_2 w \; .
\ee
The solution of these equations is explained in Appendix C and it yields
$$
w =\left ( \frac{\gm_p}{g\gm_2}\right )^2 {\rm sech}^2 \left (
\frac{t-t_0}{\tau_p}\right ) \; ,
$$
\be
\label{93}
s = -\frac{\gm_p}{g\gm_2}\; {\rm tanh}\left ( \frac{t-t_0}{\tau_p}
\right ) + \frac{1+\kappa}{g} \; .
\ee
Here
\be
\label{94}
\tau_p \equiv 1/\gm_p
\ee
is the pulse time showing the duration of the coherent relaxation 
occurring as a fast pulse. The delay time
\be
\label{95}
t_0 = t_c + \frac{\tau_p}{2}\; \ln\left | \frac{\gm_p+\gm_g}{\gm_p-\gm_g}
\right |
\ee
defines the time of the maximal coherence. The pulse width is given by the 
relation
\be
\label{96}
\gm_p^2 = \frac{1}{2}\; \gm_g^2 \left [ 1 +\sqrt{1 + 4\left (
\frac{g\gm_2}{\gm_g}\right )^2\; w_c}\; \right ] \; ,
\ee
in which
\be
\label{97}
\gm_g \equiv \gm_2 (gs_c - 1 -\kappa) \; .
\ee
The boundary values $w_c$ and $s_c$ are
\be
\label{98}
w_c = w_0 + 2\left [ \gm_3 s_0^2 - \gm_2 \left ( 1  - s_0^2 +\kappa
\right ) w_0 \right ] t_c \; , \qquad s_c=s_0(1-\gm_3t_c ) \; ,
\ee
with the chaos time $t_c$ given in Eq. (87). Since we are interested in 
the self-organized collective process, when there is no large transverse 
polarization imposed on the system at the initial time, we may set $w_0\ll 
s_0^2$.  Then Eq. (96) simplifies to
\be
\label{99}
\gm_p^2 = \gm_g^2 + (g\gm_2)^2 w_c \; .
\ee
The pulse time (94) reads as
\be
\label{100}
\tau_p = \frac{T_2}{\sqrt{(gs_c-1-\kappa)^2+g^2 w_c}} \; .
\ee
It is easy to notice that if the spin-resonator coupling is weak, 
$g\ll 1$, then $\gm_p\sim\gm_g\sim\gm_2$ and $\tau_p\sim T_2$. In that 
case, no self-organized coherence can arise in the system.

Collective coherent effects appear in the spin motion only if the pulse 
time $\tau_p$ is smaller than the dephasing time $T_2$. The inequality 
$\tau_p<T_2$, according to Eq. (100), requires that
\be
\label{101}
(gs_c - 1 -\kappa)^2 + g^2 w_c > 1 \; .
\ee
Three different regimes can satisfy Eq. (101).

The regime of {\it collective induction} happens when
\be
\label{102}
gs_0 < 1 +\kappa \; , \qquad g^2 w_0 > 1 \; .
\ee
Then, as is clear from Eq. (97), one has $\gm_g<0$, because of which 
$t_0<t_c$. This means that there is no a noticeable maximum in the coherence 
function $w$, since, by definition, the delay time (95) should occur after 
the chaotic stage, so that $t_0>t_c$. But the latter implies that 
$\gm_g>0$. 

The {\it triggered coherent relaxation} corresponds to
\be
\label{103}
gs_0 > 1 + \kappa\; , \qquad 0 < g^2w_0 < 1 \; .
\ee
And the purely {\it self-organized coherent relaxation} takes place when
\be
\label{104}
gs_0 > 2 +\kappa \; , \qquad w_0 = 0 \; .
\ee
In this classification, we keep in mind the inequality $\gm_3t_c\ll 1$, 
owing to which $w_c\approx w_0$ and $s_c\approx s_0$. The initial 
coherence is assumed to be weak, so that $w_0\ll 1$.

For $w_0\ll s_0^2$, the delay time (95) can be represented as
\be
\label{105}
t_0 = t_c +\frac{\tau_p}{2}\; 
\ln\; \frac{4(gs_c-1-\kappa)^2}{g^2 w_c} \; .
\ee
In the case of the purely self-organized coherent relaxation, for 
sufficiently large coupling and initial polarization, such that $gs_0\gg 1$, 
the delay time (105) reduces to
\be
\label{106}
t_0 = t_c + \frac{\tau_p}{2} \; 
\ln\left | \frac{2}{\gm_3 t_c}\right | \; ,
\ee
where $\tau_p=T_2/gs_0$. From these formulas, one sees that if $\gm_3\ra 0$, 
then $t_0\ra\infty$, and no coherent relaxation is possible. This emphasizes
the crucial role of the local spin fluctuations, whose existence results in
the relaxation rate $\gm_3$.

At the delay time (95), solutions (93) are given by the expressions
\be
\label{107}
w(t_0) = w_c +\left ( s_c -\; \frac{1+\kappa}{g} \right )^2 \; , \qquad
s(t_0) = \frac{1+\kappa}{g} \; .
\ee
And for $t\gg t_0$, they exponentially decay to the values
$$
w\simeq 4 w(t_0)\exp(-2\gm_p t) \; , 
$$
\be
\label{108}
s \simeq -s_c + \frac{2}{g}\; (1 +\kappa) + 2\left ( s_c -\; 
\frac{1+\kappa}{g}\right ) \exp(-2\gm_p t) \; .
\ee

At very large times $t\sim T_1$, the transient equations (91) and (92) are
no longer valid. Then one has to return to the full equations (79) and (80).
With increasing time, the solutions tend to the stationary points defined 
by the zeros of the right-hand sides of these equations. Among the relaxation 
regimes to the stationary solutions, one is especially interesting, going 
through a long series of coherent pulses. This pulsing coherent relaxation 
takes place under a permanent external pumping described by a large pumping 
rate $\gm_1^*\gg\gm_1$. Then $\Gm_1=\gm_1^*$. If also the coupling parameter 
is sufficiently large, such that $g\zeta\gg 1$ and
$$
\frac{\gm_3}{g\zeta\gm_1^*} \ll 1 \; ,
$$
then the fixed point of Eqs. (79) and (80) is given by the expressions
\be
\label{109}
w^* =\frac{\gm_1^*}{g(\gm_2+\gm_2^*)}\left ( 1  -\;
\frac{\gm_3}{g\zeta\gm_1^*}\right ) \; , \qquad
s^* = \frac{1}{g}\left ( 1 -\; 
\frac{\gm_3}{g\zeta\gm_1^*}\right ) \; ,
\ee
corresponding to a stable focus. The relaxation to the stationary solutions 
(109) realizes through a series of sharp coherent pulses, similar to the form 
of Eqs. (93), with the temporal interval between the pulses asymptotically 
defined by the separation time
\be
\label{110}
T_{sep} = \frac{2\pi}{\sqrt{2g\zeta\gm_1^*(\gm_2+\gm_2^*)}} \; .
\ee
The number of the separate coherent pulses can be estimated as 
$N_{sep}=1/\gm_1^* T_{sep}$, which gives
$$
N_{sep} = \sqrt{\frac{g\zeta(\gm_2+\gm_2^*)}{2\pi^2\gm_1^*}} \; .
$$
Such a highly nontrivial relaxation regime occurs only under a strong 
pumping and a sufficiently strong coupling with a resonator.

\section{Influence of Cross Correlations}

When in the sample, in addition to the studied spins, there are spins 
of other nature, the presence of the latter can certainly influence the 
dynamics of the former. Let us consider the case of two types of coexisting 
spins, $S$ and $F$. The total Hamiltonian is the sum
\be
\label{111}
\hat H = \hat H_S +\hat H_F + \hat H_{SF} 
\ee
of the Hamiltonians for $S$-spins, $F$-spins and their interactions. The 
Hamiltonian $\hat H_S$ of $S$-spins is the same as in Eqs. (1) to (4). Let 
us accept for the Hamiltonian $\hat H_F$ of $F$-spins a similar general form
\be
\label{112}
\hat H_F =\sum_i \hat H_{iF} + \frac{1}{2}\sum_{i\neq j} \hat H_{ijF} \; .
\ee
The single-spin terms are
\be
\label{113}
\hat H_{iF} = -\mu_{0F}\bB\cdot{\bf F}_i - D_F(F_i^z)^2 \; ,
\ee
with the total magnetic field (3). And the interaction terms are given by
\be
\label{114}
\hat H_{ijF} = \sum_{\al\bt} D_{ijF}^{\al\bt} F_i^\al F_j^\bt - J_{ijF}
{\bf F}_i \cdot {\bf F}_j \; ,
\ee
with the dipolar tensor
$$
D_{ijF}^{\al\bt} = \frac{\mu_{0F}^2}{r_{ij}^3} \left ( \dlt_{\al\bt} -
3n_{ij}^\al\; n_{ij}^\bt\right ) \; .
$$
Assume that the interactions between the $S$-and $F$-spins are represented 
by the Hamiltonian
\be
\label{115}
\hat H_{SF} = \sum_i A\bS_i\cdot{\bf F}_i + \sum_{i\neq j}
\sum_{\al\bt} A_{ij}^{\al\bt} S_i^\al F_j^\bt \; ,
\ee
containing the part of the single-site interactions of intensity $A$ and 
the part of the dipole interactions, with the dipolar tensor
$$
A_{ij}^{\al\bt} = \frac{\mu_0\mu_{0F}}{r_{ij}^3}\; \left (
\dlt_{\al\bt} - 3 n_{ij}^\al\; n_{ij}^\bt \right ) \; .
$$
In particular, these could be hyperfine interactions between nuclear and 
electron spins [15,50].

We employ notation (8) for the interaction parameters of $S$-spins and an 
equivalent notation for the interaction parameters $a_{ijF}$, $b_{ijF}$, and
$c_{ijF}$ of $F$-spins. Similarly, we define the interaction parameters
\be
\label{116}
\overline a_{ij} \equiv A_{ij}^{zz} \; , \qquad \overline b_{ij} \equiv
\frac{1}{4}\left ( A_{ij}^{xx} - A_{ij}^{yy} - 2i A_{ij}^{xy}
\right ) \; , \qquad \overline c_{ij} \equiv \frac{1}{2}\left (
A_{ij}^{xz} - i A_{ij}^{yz}\right )
\ee
for the spin cross interactions.

The local fields (11), acting on $S$-spins, are generalized to the form
$$
\xi_0 \equiv \frac{1}{\hbar} \sum_{j(\neq i)}\left [ a_{ij} S_j^z +
c_{ij}^* S_j^- + c_{ij} S_j^+ + J_{ij}(S_i^z - S_j^z) + 
\overline a_{ij} F_j^z + \overline c_{ij}^* F_j^- + \overline c_{ij}F_j^+
\right ] \; ,
$$
$$
\xi \equiv \frac{i}{\hbar} \sum_{j(\neq i)}\left [ 2c_{ij} S_j^z -\;
\frac{1}{2}\;
a_{ij} S_j^- + 2b_{ij} S_j^+ + J_{ij}(S_i^- - S_j^-) + \right.
$$
\be
\label{117}
\left.
+ 2\overline c_{ij} F_j^z -\; \frac{1}{2}\; \overline a_{ij} F_j^- + 
2\overline b_{ij}F_j^+ \right ] \; .
\ee
Analogous local fields act on $F$-spins,
$$
\xi_{0F} \equiv \frac{1}{\hbar} \sum_{j(\neq i)}\left [ a_{ijF} F_j^z + 
c_{ijF}^* F_j^- + c_{ijF} F_j^+ + J_{ijF}(F_i^z - F_j^z) +
\overline a_{ij} S_j^z + \overline c_{ij}^* S_j^- + \overline c_{ij}S_j^+
\right ] \; ,
$$
$$
\xi_F \equiv \frac{i}{\hbar} \sum_{j(\neq i)}\left [ 2c_{ijF} F_j^z -\;
\frac{1}{2}\; a_{ijF} F_j^- + 2b_{ijF} F_j^+ + J_{ijF}(F_i^- - F_j^-) +
\right.
$$
\be
\label{118}
\left. + 2\overline c_{ijF} S_j^z -\; \frac{1}{2}\; \overline a_{ij} S_j^- 
+ 2\overline b_{ij} S_j^+ \right ] \; .
\ee
Instead of one effective force (12), we have now two forces
$$
f \equiv -\; \frac{i}{\hbar}\; \mu_0(B_1+H) + \frac{i}{\hbar}\; A_iF_i^-
+\xi \; , 
$$
\be
\label{119}
f_F \equiv -\; \frac{i}{\hbar}\; \mu_{0F}(B_1+H) + \frac{i}{\hbar}\; 
A_iS_i^- + \xi_F \; .
\ee
In addition to frequency (13), let us introduce the effective frequencies
\be
\label{120}
\om_{0F} \equiv -\; \frac{\mu_{0F}}{\hbar}\; B_0\; , \qquad \ep\equiv
\frac{A}{\hbar} \; .
\ee

The Heisenberg equations of motion for the system with Hamiltonian (111) 
yield the equations for $S$-spins
$$
\frac{dS_i^-}{dt} = - i\left (\om_0 +\ep F_i^z +\xi_0\right ) S_i^- 
+ S_I^z f + \frac{i}{\hbar}\; D \left ( S_i^- S_i^z + S_i^z S_i^-
\right ) \; ,
$$
\be
\label{121}
\frac{dS_i^z}{dt} = -\; \frac{1}{2}\left ( f^+ S_i^- + S_i^+ f\right )\; ,
\ee
and the equations for $F$-spins
$$
\frac{dF_i^-}{dt} = - i\left (\om_{0F} +\ep S_i^z +\xi_{0F}\right ) 
F_i^- + F_I^z f_F + \frac{i}{\hbar}\; D_F \left ( F_i^- F_i^z + F_i^z 
F_i^- \right ) \; ,
$$
\be
\label{122}
\frac{dF_i^z}{dt} = -\; \frac{1}{2}\left ( f_F^+ F_i^- + F_i^+ f_F
\right )\; .
\ee
Again we assume that the sample is inserted into the coil of a resonant 
electric circuit. The feedback field acting on the sample is given by Eq. 
(53) or (55), where now the magnetic-moment density is
\be
\label{123}
m_x = \frac{\mu_0}{V}\; \sum_{i=1}^N < S_i^x> \; + \;
\frac{\mu_{0F}}{V}\; \sum_{j=1}^{N_F} < F_j^x>\; ,
\ee
with $N_F$ being the number of $F$-spins.

Averaging Eqs. (121) and (122), we derive the evolution equations for 
functions (29), (30), and (31), corresponding to $S$-spins, as well as the
equations for the functions
\be
\label{124}
u_F \equiv \frac{1}{FN_F} \; \sum_{i=1}^{N_F} <F_i^-> \; ,
\ee
\be
\label{125}
w_F \equiv \frac{1}{F^2N_F(N_F-1)} \; \sum_{i\neq j}^{N_F} 
<F_i^+F_j^-> \; ,
\ee
\be
\label{126}
s_F \equiv \frac{1}{FN_F} \; \sum_{i=1}^{N_F} <F_i^z> \; ,
\ee
describing $F$-spins. In this notation, the transverse magnetic-moment 
density (123) is
$$
m_x = \frac{1}{2}\; \rho\mu_0 S(u^*+u) + \frac{1}{2}\; \rho_F \mu_{0F}
F(u_F^* + u_F ) \; ,
$$
where $\rho_F$ is the density of $F$-spins.

The analysis of the evolution equations for the combined system of $S$- 
and $F$-spins is the same as has been given above for one type of spins 
$S$, with the difference that all expressions become much more cumbersome. 
Again it is possible to show that in the triggering of spin motion an 
important role is played by the coupled $S-F$ spin fluctuations, which 
yield the dynamic relaxation rates $\gm_3$ and $\gm_{3F}$ defined by 
the relations
\be
\label{127}
\gm_3^2 = \gm_{SS}^2 + \gm_{SF}^2\; , \qquad \gm_{3F}^2 = \gm_{FF}^2 + 
\gm_{FS}^2\; ,
\ee
where
$$
\gm_{SS} \approx \rho\; \frac{\mu_0^2}{\hbar}\; \sqrt{S(S+1)} \; , \qquad 
\gm_{SF} \approx \sqrt{\rho\rho_F}\; \frac{\mu_0\mu_{0F}}{\hbar}\; F \; ,
$$
$$
\gm_{FF} \approx \rho_F\; \frac{\mu_{0F}^2}{\hbar}\; \sqrt{F(F+1)} \; , 
\qquad \gm_{FS} \approx \sqrt{\rho\rho_F}\; 
\frac{\mu_{0F}\mu_0}{\hbar}\; S \; .
$$

The effective frequencies of $S$- and $F$-spins, respectively, are
\be
\label{128}
\om_S =\om_0 -\om_D s + \ep s_F S\; , \qquad
\om_F =\om_{0F} -\om_{DF} s_F + \ep s F\; ,
\ee
where $\om_D$ is given by Eq. (19) and
\be
\label{129}
\om_{DF} \equiv (2F-1)\; \frac{D_F}{\hbar} \; .
\ee

We shall not overload this paper by a detailed exposition of various cross 
correlations resulting from the complicated system of the coupled evolution 
equations for $S$- and $F$-spins. Let us only emphasize the existence of 
a rather nontrivial nonlinear effect of mutual spin interactions through 
the resonator feedback field. Calculating the latter from the integral 
representation (55), with the transverse magnetic density (123), and 
substituting this into the evolution equations results in an effective 
mutual influence of spins through the feedback field. If the resonator is 
tuned to the characteristic frequency $\om_S$ of $S$-spins, then for the 
latter, we derive the evolution equations similar to Eqs. (79) and (80), 
but with the effective spin-resonator coupling
\be
\label{130}
g =\frac{\gm\gm_0\om_S}{\gm_2(\gm^2+\Dlt^2)} \left ( 1 + 
\frac{\rho_F\mu_{0F}\ep s_F F}{\rho\mu_0\om_F} \right ) \; ,
\ee
instead of Eq. (63), and with $\gm_3$ given by Eq. (127). Depending on the 
spin characteristics, coupling (130) can substantially surpass the value of 
Eq. (63). This is because the subsystem of $F$-spins, coupled to a resonator, 
becomes itself a kind of an additional resonator for $S$-spins.

\section{Conclusion}

A general theory is developed for describing nonlinear spin relaxation, 
which occurs when the spin system is prepared in a strongly nonequilibrium 
state and when the sample is coupled to a resonator electric circuit. A 
strongly nonequilibrium initial state can be realized by placing a polarized 
sample into an external magnetic field, whose direction is opposite to the 
sample magnetization.  Nonlinearity in spin relaxation comes from direct 
spin-spin interactions and from their effective interactions through the 
resonator feedback field. Direct spin interactions are responsible for the 
appearance of local spin fluctuations, playing a crucial role at the 
starting stage of relaxation. The resonator feedback field collectivizes 
the spin motion, leading to coherent collective relaxation. The developed 
theory is based on a realistic Hamiltonian containing  the main spin 
interactions. The role of various relaxation rates is thoroughly analysed.

The aim of the present paper has been to develop a general theory providing
an accurate and realistic description of nonlinear spin relaxation. This 
theory can be employed for a large class of polarized spin materials. 
Applications to particular substances require a special consideration and
separate publications. There exists a large variety of materials that can 
be treated by the developed theory. Just to give an example, we may 
mention the class of molecular magnets [16,19,24--26]. For instance, the 
molecular crystal V$_{15}$ is made of molecules of spin $1/2$, so has no 
magnetic anisotropy. Its nonlinear spin relaxation can be realized in a 
rather weak external field $B_0\geq 1$ G. The molecules Mn$_{12}$ and Fe$_8$ 
possess the spin $S=10$. They form crystals with density $\rho\sim 10^{21}$ 
cm$^{-3}$. The anisotropy frequency is $\om_D\sim 10^{12}$ s$^{-1}$. At 
low temperatures below about 1 K, the molecules can be well polarized, 
with the spin-lattice relaxation parameters $\gm_1\sim 10^{-5}-10^{-7}$ 
s$^{-1}$. The line width is caused by rather strong dipole interactions, 
with $\gm_2\sim 10^{10}$ s$^{-1}$. The condition $\om_0>\om_D$ can be 
reached for $B_0>10^5$ G. In the molecular magnet, formed by the molecules 
Mn$_6$, whose spin is $S=12$, the magnetic anisotropy is much weaker, with 
$\om_D\sim 10^{10}$ s$^{-1}$, being of the same order as $\gm_2\sim 10^{10}$ 
s$^{-1}$. Therefore the required magnetic field is not high, $B_0> 10^3$ G.
Coupling a molecular crystal to a resonant circuit with the natural width 
$\gm\equiv\om/2Q$, where $Q$ is the resonator quality factor, one can attain 
the values of the coupling parameter as large as $g\sim Q\sim 10^4$. With 
such a strong coupling, the influence of the resonator feedback field 
outperforms other relaxation mechanisms, producing fast coherent relaxation, 
with relaxation times $\tau_p\sim 10^{-13}$ s. Such a fast reorientation 
of the magnetic moment can result in the emission of radiation pulses of 
high intensity.

\vskip 5mm

{\bf Acknowledgements}

\vskip 2mm

I am grateful to E.P. Yukalova for helpful discussions. I appreciate the 
Mercator Professorship of the German Research Foundation.

\newpage

{\Large{\bf Appendix A: Effective Homogeneous Broadening}}

\vskip 5mm

The homogeneous broadening, existing in spin systems, arises from spin-spin 
interactions and is usually expressed through the moments $M_n$, which may 
depend on the level of the longitudinal polarization $s$, provided the latter 
is sufficiently large. The moments have been calculated in a number of works 
[1--7,21]. The most general and exact formula, relating the effective 
broadening with the moments, can be found in Abragam and Goldman [6], 
which for the Gaussian line shape is
$$
\gm_2(s) =\sqrt{\frac{\pi M_2^3(s)}{2[M_4(s)-M_2^2(s)]} } \; .
$$
The Lorentzian line shape yields to practically the same expression, with a 
slightly different coefficient. The broadening $\gm_2(s)$ for the Lorentzian 
line is $\sqrt{\pi}$ of the Gaussian broadening. The dependence of the moments
on the polarization has been accurately calculated [6], yielding
$$
M_2(s) =M_2(0)(1-s^2) \; , \qquad
M_4(s) = 2.18 M_2^2(0)(1-s^2)(1-0.42 s^2) \; .
$$
Substituting this into $\gm_2(s)$, and taking into account that $s^2\leq 1$, 
we obtain Eq. (34).

\newpage

{\Large{\bf Appendix B: Spin Radiation Rate}}

\vskip 5mm

To get a fully quantum-mechanical microscopic picture of spin interactions 
with electromagnetic field they radiate, one has to add to the spin 
Hamiltonian (1) the field Hamiltonian
$$
\hat H_f = \frac{1}{8\pi} \; \int \left ( {\bf E}^2 +{\bf H}^2 \right ) \;
d\br \; ,
$$
where ${\bf E}={\bf E}(\br,t)$ is electric field and ${\bf H}={\bf H}(\br,t)$ 
is magnetic field, and the operator energy of spin-field interactions
$$
\hat H_{sf} = -\mu_0 \sum_{i=1}^N {\bf S}_i\cdot{\bf H}_i \; ,
$$
where ${\bf H}_i={\bf H}(\br_i,t)$. From the Heisenberg equations of 
motion for the field variables, one finds the vector potential
$$
{\bf A}(\br,t) = \frac{1}{c}\; \int {\bf j}\left (\br', t -\;
\frac{|\br-\br'|}{c}\right )\; \frac{d\br}{|\br-\br'|} \; ,
$$
in which the current density is
$$
{\bf j} = -c\mu_0 \sum_{i=1}^N 
{\bf S}_i\times\vec\nabla \dlt(\br -\br_i) \; .
$$
The vector potential ${\bf A}_i\equiv{\bf A}(\br_i,t)$ can be represented 
as
$$
{\bf A}_i = {\bf A}_i^- + {\bf A}_i^+ + {\bf A}_i' \; ,
$$
where
$$
{\bf A}_i^- = -\sum_j \left ( 1 + \frac{1}{c}\; \frac{\prt}{\prt t}\right )
\frac{\br_{ij}}{r_{ij}^3} \times \vec\mu^*\; S_j^-\left ( t -\;
\frac{r_{ij}}{c}\right ) \; ,
$$
$$
{\bf A}_i' = -\sum_j \; \frac{\br_{ij}}{r_{ij}} \times\vec\mu_0\; S_j^z 
\left ( t - \; \frac{r_{ij}}{c}\right ) \; ,
$$
with the notation
$$
\vec\mu \equiv \frac{\mu_0}{2}({\bf e}_x  - i{\bf e}_y ) \; , \qquad
\vec\mu_0 \equiv\mu_0{\bf e}_z \; .
$$
From here, we get the magnetic field ${\bf H}_i\equiv{\bf H}(\br_i,t)$ 
acting on an $i$-th spin as ${\bf H}_i=\vec\nabla_i\times{\bf A}_i$, which 
gives the field
$$
{\bf H}_i = {\bf H}_i^- + {\bf H}_i^+ + {\bf H}_i' \; ,
$$
in which
$$
{\bf H}_i^-  = -\sum_j \left [ 
\frac{\vec\mu^*-(\vec\mu^*\cdot{\bf n}_{ij}){\bf n}_{ij}}{c^2r_{ij}}\; 
\frac{\prt^2}{\prt t^2} + 
\frac{\vec\mu^*-3(\vec\mu^*\cdot{\bf n}_{ij}){\bf n}_{ij}}{r_{ij}^3}\;
\left ( 1 +\frac{r_{ij}}{c}\; \frac{\prt}{\prt t}\right ) \right ]
S_j^-\left ( t  -\; \frac{r_{ij}}{c}\right ) \; ,
$$
$$
{\bf H}_i' = -\sum_j 
\frac{\vec\mu_0-3(\vec\mu_0\cdot{\bf n}_{ij}){\bf n}_{ij}}{r_{ij}^3}\;
S_j^z \left ( t  -\; \frac{r_{ij}}{c}\right ) \; .
$$
If the spins on different sites move independently of each other, so that 
the single-spin terms in the above sums chaotically oscillate, then the 
average magnetic field acting on each spin from the radiation of other 
spins is zero. Noticeable action of other spins can arise only if there 
exist the groups of spins, the so-called spin packets, which are strongly 
correlated, moving together. A substantial mutual interaction between spins, 
caused by their electromagnetic radiation, can appear only when this 
radiation is monochromatic, with a well-defined spin frequency $\om_s$, the
related wavelength $\lbd=2\pi c/\om_s$, and wave vector $k=\om_s/c$. This 
radiation can collectivize spins in a spin packet of size $L_s$, provided 
that
$$
k L_s \ll 1 \; .
$$
When the radiation wavelength $\lbd$ is much larger than the system length 
$L$, then $L_s=L$. This, however, is not compulsory, and the size of a spin
packet can be much shorter than $L$, but it should be much larger than the 
mean interspin distance. Thus, inequality (38) is a necessary condition 
for the appearance of collective effects.

Under condition (38), the above magnetic fields can be simplified, 
averaging them over spherical angles. The resulting expressions have to be 
added to the magnetic field in the effective force (12), which acquires 
one more term, being the friction force
$$
f' = ( \gm_r - i\dlt\om ) u \; ,
$$
in which the collective radiation rate and frequency shift are
$$
\gm_r \equiv \gm_0 \sum_j^{N_s} \frac{\sin(kr_{ij})}{kr_{ij}}\;
\Theta(ct-r_{ij}) \; ,
$$
$$
\dlt\om \equiv \gm_0 \sum_j^{N_s} \frac{\cos(kr_{ij})}{kr_{ij}}\;
\Theta(ct-r_{ij}) \; ,
$$
where
$$
\gm_0 \equiv \frac{2}{3\hbar}\; \mu_0^2 S k^3 
$$
is the single-spin natural width, $\Theta(\cdot)$ is a unit-step function, 
and $N_s=\rho L_s^3$ is the number of spins in a spin packet. These 
formulas can be further simplified to 
$$
\gm_r = \gm_0 N_s = \frac{2}{3\hbar}\; \mu_0^2 S k^3 N_s
$$
and
$$
\dlt\om = \frac{3\gm_r}{2kL_s} = \frac{1}{\hbar}\; 
\rho\mu_0^2 S(kL_s)^2  \; .
$$
The frequency shift is very small, even as compared to $\gm_2$, since
$$
\frac{\dlt\om}{\gm_2} \cong 0.1 (kL_s)^2 \ll 1 \; .
$$
Of course, such a small shift can be omitted, being negligible as compared 
to $\gm_2$ and the more so as compared to $\om_s$. And for the radiation 
rate $\gm_r$, substituting there $N_s=\rho L_s^3$, we obtain Eq. (37).

\newpage

{\Large{\bf Appendix C: Transient Stage of Relaxation}}

\vskip 5mm

After the chaotic stage of spin fluctuations, the transient stage comes 
into play, characterized by Eqs. (91) and (92). The latter, by introducing 
the function
$$
y \equiv \gm_2\left (1 - s^2 + \kappa - gs\right )
$$
and keeping in mind a sufficiently large coupling parameter $g\gg s$, 
rearrange to
$$
\frac{dw}{dt} = - 2y w\; , \qquad \frac{dy}{dt} =(g\gm_2)^2 w \; .
$$
Differentiating the second of these equations, we have
$$
\frac{d^2y}{dt^2}  + 2y\; \frac{dy}{dt} = 0 \; ,
$$
which yields
$$
\frac{dy}{dt} + y^2 = \gm_p^2 \; ,
$$
with $\gm_p$ being an integration parameter. This Riccati equation possesses
the solution
$$
y = \gm_p{\rm tanh}\left ( \frac{t-t_0}{\tau_p} \right ) \; ,
$$
in which $\gm_p\tau_p\equiv 1$ and $t_0$ is another integration constant. 
Inverting the dependence of $y$ on $s$ for $s^2\leq 1$, we get
$$
s = -\; \frac{y}{g\gm_2} + \frac{1+\kappa}{g} \; .
$$
This gives the second of Eqs. (93), while the first of solutions (93) 
follows from Eq. (92). The integration constants $\gm_p$ and $t_0$ are 
defined by the initial conditions, which for the transient stage are 
$w_c=w(t_c)$ and $s_c=s(t_c)$.

\end{document}